\documentclass[preprint,showpacs, superscriptaddress, preprintnumbers,amsmath,amssymb]{revtex4}

\usepackage{booktabs}
\usepackage{color,graphicx}
\usepackage{amsmath}
\usepackage{dcolumn}
\usepackage{CJK}                 
\usepackage{bm}                  
\usepackage{soul}
\usepackage{enumerate}
\usepackage{lmodern}
\usepackage[driverfallback=dvipdfm,bookmarks=true,colorlinks,%
           citecolor=blue,linkcolor=blue,%
           breaklinks=true]{hyperref}

\begin{document}

\title{Stability of the linear chain structure for $^{12}$C in covariant density functional theory on a 3D lattice}
\author{Z. X. Ren}
\affiliation{State Key Laboratory of Nuclear Physics and Technology, School of Physics, Peking University, Beijing 100871, China}

\author{S. Q. Zhang}
\affiliation{State Key Laboratory of Nuclear Physics and Technology, School of Physics, Peking University, Beijing 100871, China}

\author{P. W. Zhao}
\affiliation{State Key Laboratory of Nuclear Physics and Technology, School of Physics, Peking University, Beijing 100871, China}

\author{N. Itagaki}
\affiliation{Yukawa Institute for Theoretical Physics, Kyoto University, Kyoto 606-8502, Japan}

\author{J. A. Maruhn}
\affiliation{Institut f\"{u}r Theoretische Physik, Goethe-Universit\"{a}t, Max-von-Laue-Str. 1, 60438 Frankfurt am Main, Germany}

\author{J. Meng}
\email{mengj@pku.edu.cn}
\affiliation{State Key Laboratory of Nuclear Physics and Technology, School of Physics, Peking University, Beijing 100871, China}
\affiliation{Yukawa Institute for Theoretical Physics, Kyoto University, Kyoto 606-8502, Japan}
\affiliation{Department of Physics, University of Stellenbosch, Stellenbosch, South Africa}


\date{\today}

\begin{abstract}
  The stability of the linear chain structure of three $\alpha$ clusters for $^{12}$C against the bending and fission is investigated in the cranking covariant density functional theory,
  in which the equation of motion is solved on a 3D lattice with the inverse Hamiltonian and the Fourier spectral methods.
  Starting from a twisted three $\alpha$ initial configuration, it is found that the linear chain structure is stable when the rotational frequency is within the range of $\sim$2.0 MeV to $\sim$2.5 MeV.
  Beyond this range, the final states are not stable against fission.
  By examining the density distributions and the occupation of single-particle levels, however, these fissions are found to arise from the occupation of unphysical continuum with large angular momenta.
  To properly remove these unphysical continuum, a damping function for the cranking term is introduced.
  Eventually, the stable linear chain structure could survive up to the rotational frequency $\sim$3.5 MeV, but the fission still occurs when the rotational frequency approaches to $\sim$4.0 MeV.

\end{abstract}


\maketitle

\section{INTRODUCTION}
The nuclear deformation reflects the anisotropic mass distribution viewed from the intrinsic coordinate frame of nuclei \cite{Bohr&Mottelson1975}.
The deformation can be identified by the features of the observed excitation spectra.
In heavy nuclei, evidences for the deformation with length-to-width ratios of 2:1 or 3:1 have been provided by the so-called superdeformed \cite{NyakoPRL1984Superdeform, TwinPRL1986Superdeform} and hyperdeformed bands \cite{Galindo-Uribarri1993hyperdeform, LaFosse1995hyperdeform, Krasznahorkay1998hyperdeform}.
In light nuclei, more exotic states, such as the linear chain states (LCSs), might exist due to the $\alpha$ clustering.

A typical example of the LCSs was suggested in $^{12}$C about 60 years ago \cite{Morinaga1956LinearChain} for the structure of the Hoyle state (the second $0^+$ state with excitation energy $E_x=7.65$ MeV).
However, later investigation indicates that the Hoyle state is a gas-like state rather than a state with geometrical configurations \cite{Fujiwara1980AlphaNuclei} and
reinterpreted as an $\alpha$-condensate-like state in some recent works \cite{Tohsaki2001AlphaClusterC12_O16, Suhara2014ClusterCondensation}.
Since then, various theoretical and experimental works have been done to search for the LCSs in not only $^{12}$C but also other nuclei,
such as C isotopes \cite{Zhao2015Rod-shaped, Itagaki2001MolecularOrbit, Milin2002, Oertzenn2004C14_LCS, Itagaki2006C13_LCS, Suhara2010C14ClusterExcited, Furutachi2011BendingLCSC13, Freer2014C14_LCS_C14, Baba2014LCS_C16, He2014C12O16_CLuster, Liu2012CPC_C12O16Cluster, He2016Cluster, Fritsch2016C14LCS, Li2017C14LCS, Yamaguchi2017C14LCS, Baba2017C14_LCS_decay},
$^{16}$O \cite{Yao2014searching, Chevallier1967O16LCS, Suzuki1972alphaChainDecay, Flocard1984PTEP_OCluster, Bender2003NPAcluster, Ichikawa2011O16LinearChain},
and $^{24}$Mg \cite{Wuosmaa1992Mg24LCS,Iwata2015PRC_LCS_Mg}.
For $^{12}$C, the existence of bending motion in three $\alpha$ chain around the $E_x = 10$ MeV region has been discussed \cite{Neff2004LCS_C12, KanadaEnyo2007LCS_C12}.

The theoretical studies of LCSs have been mostly performed in conventional cluster models with effective interactions determined by the binding energies and scattering phase shifts of the clusters \cite{Oertzen2006PR_cluster, freer2017microscopic}.
Recently, more and more other approaches, such as density functional theories (DFTs), have been used to investigate the LCSs \cite{Maruhn2010NPA_LCS_Carbon, Ichikawa2011O16LinearChain, Ebran2014ClusterCDFT, Zhao2015Rod-shaped, Iwata2015PRC_LCS_Mg, Yao2014searching}.
The DFTs are designed to describe various properties of nuclei in the whole nuclear chart.
Moreover, the existence of $\alpha$ clusters is not assumed a priori in the DFTs.
Therefore, it would provide more confidence for the prediction of LCSs.
For the stabilization of the LCSs, giving angular momentum to the systems is a useful prescription,
because the centrifugal force makes these largely elongated shapes energetically favored.
For this purpose, cranking model is often utilized.
It has been discussed in $^{16}$O and $^{24}$Mg that LCSs exist, as energy minima, in a range of the rotational frequencies.
At a lower frequency, it results in other normal configurations as the favored configurations,
while a higher one leads to the fission \cite{Ichikawa2011O16LinearChain, Iwata2015PRC_LCS_Mg}.

Up to now, both nonrelativistic and relativistic DFTs have been applied to investigate the LCSs in, for examples,
C isotopes \cite{Maruhn2010NPA_LCS_Carbon, Zhao2015Rod-shaped}, $^{16}$O \cite{Ichikawa2011O16LinearChain, Yao2014searching},
and other light $N=Z$ nuclei \cite{Ebran2014ClusterCDFT, Iwata2015PRC_LCS_Mg}.
For a full understanding of the linear chain structure, in particular, for its stability against bending motion,
a three-dimensional (3D) lattice solution of the DFTs is very important.
In the 3D lattice calculation, there is no symmetry limitation for the single-particle wave functions in space,
and a much more precise description can be achieved.
This has been realized in the nonrelativistic DFTs \cite{Maruhn2010NPA_LCS_Carbon, Ichikawa2011O16LinearChain, Iwata2015PRC_LCS_Mg},
where the Schr\"odinger equations for nucleons are solved in a 3D lattice with the imaginary time method (ITM) \cite{davies1980imaginary} or the damped-gradient iteration method \cite{reinhard1982comparative}.

The relativistic DFT, namely covariant DFT (CDFT), has many advantages in describing nuclear systems,
such as the natural inclusion of the spin degree of freedom and the spin-orbit potential \cite{RING1996PPNP, Vretenar2005PhysicsReport, PPNP2006},
the interpretation of the pseudospin symmetries of nucleons and spin symmetries of antinucleons \cite{liang2015hidden},
and the self-consistent treatment of the time-odd field \cite{Afanasjev1999PhysicsReport, Meng2013FT_TAC}, see also Ref. \cite{meng2016relativistic} for details.
The CDFT has been widely applied to investigate the ground states of nuclei and various rotational excitation phenomena,
including magnetic \cite{Peng2008maganetic_roration, Zhao2011Ni60}
and antimagnetic rotation \cite{Zhao2011PRL_AMR, zhao2012PRC_AMR}, multiple chiral doublet bands \cite{Meng2006MxD, Meng2016CDFT_chiral_MR, Zhao2017ChiralRotation},
and the rotation of LCSs \cite{Yao2014searching, Zhao2015Rod-shaped}.
In particular, in Ref. \cite{Zhao2015Rod-shaped}, it is found a strong hint that LCSs could be realized in nuclei with extreme spin and isospin.

To have a full understanding of the LCSs, the relativistic study based on the 3D lattice method is required.
Different from the Schr\"odinger equation, a direct implementation of the ITM to solve the Dirac equation suffers several serious problems,
including the variational collapse \cite{ZhangIJMPE2010} and Fermion doubling problems \cite{wilson1977new, tanimura20153d}.
To avoid the variational collapse problem, Hagino and Tanimura adopted the idea of an inverse Hamiltonian proposed by Hill and Krauthauser \cite{hill1994solution} and solved the spherical Dirac equation with ITM \cite{hagino2010iterative}.
However, when they extended this method to the Dirac equation in 3D lattice space,
the Fermion doubling problem appears due to the replacement of the derivative by the finite-difference method \cite{tanimura20153d}.
Recently, this Fermion doubling problem has been solved by adopting the Fourier spectral method \cite{REN2017Dirac3D}.

In this work, the equation of motion in cranking covariant density functional theory is solved on a 3D lattice
with the inverse Hamiltonian and the Fourier spectral methods.
The CDFT in 3D lattice space is then applied to study the stability of the LCSs in $^{12}$C.
The theoretical framework will be briefly introduced in Sec. \ref{TheoreticalFrame}.
In Sec. \ref{NumDetail}, the numerical details are presented.
Sec. \ref{ResultDiscussion} is devoted to the results and discussion.
A summary is given in Sec. \ref{summary}.

\section{THEORETICAL FRAMEWORK}\label{TheoreticalFrame}
The starting point of covariant density functional theory is a standard effective Lagrangian density,
where nucleons can be coupled with either finite-range meson fields \cite{Long2004PK1, lalazissis2005new}
or zero-range point-coupling interactions \cite{Burvenich2002PCF1, Niksic2008DDPC1, ZhaoPC-PK1}.
For nuclear rotations, one can transform the effective Lagrangian into a rotating frame with a constant rotational frequency $\omega$ around a certain direction.
The optimal solution of the rotating nucleus is determined by minimizing the Routhian of the total system.
This gives rise to either the principal axis cranking CDFT \cite{Koepf1989PAC_CDFT, Koepf1990PAC, Konig1993Identicalbands}, where the cranking axis is one of the three principal axes of a nucleus,
or the tilted axis cranking one with the cranking axis different from any of the principal axes,
including planar \cite{Peng2008maganetic_roration, Zhao2011Ni60, Meng2013FT_TAC} and aplanar rotation versions \cite{Zhao2017ChiralRotation, Madokoro2000Rb84}.

The equation of motion for nucleons has the form of a Dirac equation,
\begin{equation}\label{Dirac_equation}
  \begin{split}
   \hat{h}'\psi_k=\left(\hat{h}_0-\omega\hat{j}_y\right)\psi_k=\varepsilon'_k\psi_k.
  \end{split}
\end{equation}
Here $\hat{h}'$ is the cranking single-particle Hamiltonian, and $-\omega\hat{j}_y$ is the Coriolis or cranking term.
The cranking axis is fixed as the $y$ axis, and $\hat{j}_y=\hat{l}_y+\frac{1}{2}\hat{\Sigma}_y$ is the $y$ component of the total angular momentum of the nucleon spinors.
The single-particle Hamiltonian $\hat{h}_0$ reads
\begin{equation}\label{spHamiltonian}
   \hat{h}_0=\bm{\alpha\cdot}[-i\bm{\nabla}-\bm{V}(\bm{r})]+\beta[m_N+S(\bm{r})]+V_0(\bm{r}).
\end{equation}
The $\varepsilon'_k$ represents the single-particle Routhians.
The single-particle energies are obtained by calculating the expectation values of $\hat{h}_0$ with respect to single-particle wave functions $\psi_k$.
The relativistic scalar $S(\bm{r})$ and vector $V_\mu(\bm{r})$ fields are connected in a self-consistent way to the densities and current distributions of the nucleons.
By solving the Dirac equation Eq. \eqref{Dirac_equation} self-consistently,
one can proceed to calculate various physical observables for the nuclear system,
such as angular momenta and total energies.
For the detailed formalism, one can read, for examples, Refs. \cite{Meng2013FT_TAC, meng2016relativistic}.

So far, the cranking Dirac equation is solved only in the harmonic oscillator basis \cite{Koepf1989PAC_CDFT, Zhao2011PRL_AMR, Zhao2017ChiralRotation}.
In this work, Eq. \eqref{Dirac_equation} is solved in 3D lattice space.
Similar to the solution of the static Dirac equation in Ref. \cite{REN2017Dirac3D},
the variational collapse and the Fermion doubling problems are respectively solved by the inverse Hamiltonian \cite{hagino2010iterative} and the Fourier spectral methods \cite{Shen2011Spectral}.
The wave functions are obtained by imaginary time evolution,
\begin{equation}\label{IHM_evolution}
   \psi^{(n+1)}_k=\mathcal{O}\left\{\left(1+\frac{\Delta \tau}{\hat{h}'-W_k}\right)\psi^{(n)}_k\right\},
\end{equation}
where $\mathcal{O}$ means the orthonormalization of the wave functions,
the upper indices of the wave functions indicate the iteration number,
$\Delta \tau$ is the imaginary time step, and $W_k$ is the energy shift parameter.
Here, the orthonormalization is realized by the Gram-Schmidt method.

The full space is discretized by an even number of grid points along the $x$, $y$ and $z$ axes,
and the grid points are distributed in a symmetric way around the origin point.
Taking the $x$ direction as an example, the coordinates of these grid points are arranged as,
\begin{equation}\label{spatialgrides}
     x_\nu=\left(-\frac{n_x-1}{2}+\nu-1\right)dx,\quad \nu=1,...,n_x,
\end{equation}
where $dx$ is the step size and $n_x$ is the grid number in the $x$ direction.
The spatial derivative is calculated in momentum space by the Fourier spectral method.
In the following, this method is illustrated in the one-dimensional (1D) case and it is straightforward to generalize to the 3D case.
The grid points \{$k_\mu$\} in the momentum space are related to the spatial grid points in Eq. \eqref{spatialgrides} with the equation,
\begin{equation}
    k_\mu=
    \begin{cases}
      (\mu-1)dk,&\mu=1,...,n_x/2,\\
      (\mu-n_x-1)dk,&\mu=n_x/2+1,...,n_x,
    \end{cases}
\end{equation}
where the steps are defined as $dk=2\pi/(n_x\cdot dx)$.
A given function $f(x_\nu)$ can be connected with its Fourier transform $\tilde{f}(k_\mu)$ via
\begin{subequations}
  \begin{align}
    &\tilde{f}(k_\mu)=\sum_{\nu=1}^{n_x}\exp(-\textrm{i} k_\mu x_\nu)f(x_\nu),\label{Eq_DFT}\\
    &f(x_\nu)=\frac{1}{n_x}\sum_{\mu=1}^{n_x}\exp(\textrm{i} k_\mu x_\nu)\tilde{f}(k_\mu).\label{Eq_inversDFT}
  \end{align}
\end{subequations}
From Eq. \eqref{Eq_inversDFT}, the $m$-th order derivative of $f(x_\nu)$ can be found as,
\begin{equation}
  \begin{split}
     f^{(m)}(x_\nu)&=\frac{1}{n_x}\sum_{\mu=1}^{n_x}\exp(\textrm{i} k_\mu x_\nu)(\textrm{i} k_\mu)^m\tilde{f}(k_\mu)\\
                   &=\frac{1}{n_x}\sum_{\mu=1}^{n_x}\exp(\textrm{i} k_\mu x_\nu) \widetilde{f^{(m)}}(k_\mu).
  \end{split}
\end{equation}
Here $\widetilde{f^{(m)}}(k_\mu)$ is the Fourier transform of $f^{(m)}(x_\nu)$,
\begin{equation}\label{Eq_f&fm}
  \widetilde{f^{(m)}}(k_\mu)=(\textrm{i} k_\mu)^m\tilde{f}(k_\mu).
\end{equation}
Then one could get the $m$-th order derivative of $f(x_\nu)$ by performing the inverse Fourier transform on $\widetilde{f^{(m)}}(k_\mu)$.
In the calculation, the Fourier and the inverse Fourier transforms are performed by the fast Fourier transform (FFT) technique.

\section{NUMERICAL DETAILS}\label{NumDetail}
The successful density functional DD-ME2 \cite{lalazissis2005new} is employed.
The Dirac spinors of the nucleons and the potentials in the single-particle Hamiltonian \eqref{spHamiltonian} are represented in 3D lattice space.
The step sizes along the $x$, $y$, and $z$ axes are identical and chosen as 0.8 fm.
The grid numbers are 24 for the $x$ and $y$ axes and 34 for the $z$ axis.
It turns out that the size of the space adopted here is sufficient to obtain converged solutions \cite{Ichikawa2011O16LinearChain}.
The imaginary time-step size $\Delta \tau$ is taken as 150 MeV.
As mentioned in Ref. \cite{Zhao2015Rod-shaped}, since the density of the single-particle
levels is rather low, the pairing correlations could be neglected safely.
The convergence of the iteration is achieved by requiring that the energy uncertainty for every occupied single-particle state is smaller than $10^{-9}$ MeV$^2$.

Two numerical tricks are employed to speed up the convergence of the iterations:
\begin{itemize}
  \item[(1)] The energy shift $W_k$  in Eq. \eqref{IHM_evolution} for the $k$-th level is taken as
  \begin{equation}
    W_k=\varepsilon_k'-\Delta W_k,
  \end{equation}
  where $\varepsilon_k'=\langle\psi_k|\hat{h}'|\psi_k\rangle$ with $\hat{h}'$ being cranking Dirac Hamiltonian.
  The choice of $\Delta W_k$ is as follows:
  \begin{equation}
    \Delta W_k=\begin{cases}
                  6~{\rm MeV},&k=1;\\
                  \varepsilon_k'-\varepsilon_{k-1}',&k>1~{\rm and}~\varepsilon_k'-\varepsilon_{k-1}'>\Delta W_1;\\
                  \Delta W_{k-1},&k>1~{\rm and}~\varepsilon_k'-\varepsilon'_{k-1}\leq\Delta W_1.
                \end{cases}
  \end{equation}
  \item[(2)] During the imaginary time evolution in Eq. \eqref{IHM_evolution}, the wave functions $\{\psi^{(n+1)}_k\}$ at every iteration constitute an orthonormal space.
      Similar to Ref. \cite{REN2017Dirac3D}, the cranking Dirac Hamiltonian is diagonalized within this orthonormal space at every iteration,
      and the obtained eigenfunctions are taken as the initial wave functions for the next iteration.
\end{itemize}

\section{RESULTS AND DISCUSSION}\label{ResultDiscussion}

\begin{figure}[h!]
  \centering
  \includegraphics[width=0.5\textwidth]{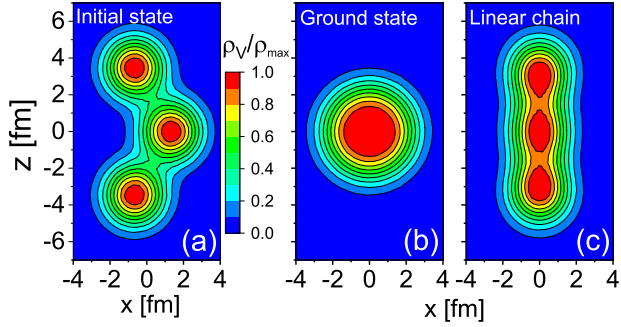}\\
  \caption{The total nucleon density distributions in the $x$-$z$ plane ($y$ direction is integrated) for (a) the initial state,
  (b) the ground state, and (c) the three-$\alpha$ LCS at rotational frequency $\hbar\omega=2.0$ MeV.
  In each figure, the density is normalized with respect to its corresponding maximum density $\rho_{\rm max}$. }\label{Fig1_DensPlane}
\end{figure}
We investigate the stability of LCSs in $^{12}$C against bending and fission.
The initial state of the cranking CDFT calculations is shown in Fig. \ref{Fig1_DensPlane} (a).
It is a twisted linear chain constructed by placing the three wave function sets of $^{4}$He.
For comparisons, the obtained density distributions for the ground state and LCS
($\hbar\omega=2.0$ MeV) of $^{12}$C are shown in Figs. \ref{Fig1_DensPlane} (b) and (c), respectively.
The density in each figure is normalized with respect to its corresponding maximum density $\rho_{\rm max}$.

\begin{figure}[h!]
  \centering
  \includegraphics[width=0.5\textwidth]{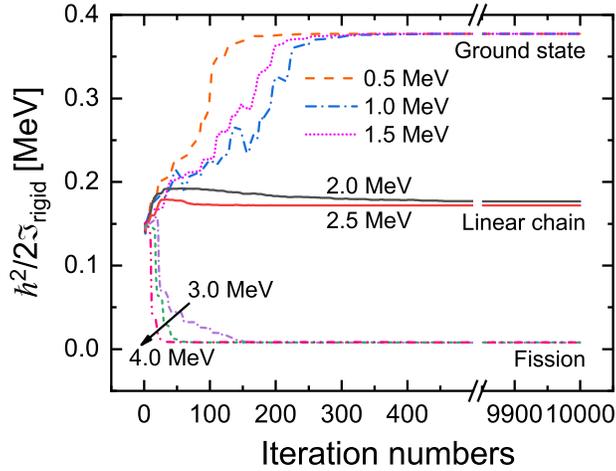}\\
  \caption{Coefficient of the rotational energy, $\hbar^2/2\Im_{\rm rigid}$, calculated by cranking CDFT as a function of iteration numbers at rotational frequencies $\hbar\omega=0.5$, 1.0, ..., 4.0 MeV.
  Ground state and LCS ($\hbar\omega=2.0$ MeV) correspond to the density distributions given in Figs.\ref{Fig1_DensPlane} (a) and (b), respectively.}\label{Fig2_MOInodam}
\end{figure}
Then we perform the self-consistent cranking CDFT calculations starting from the initial state in Fig. \ref{Fig1_DensPlane} (a).
To check the convergence of the imaginary time evolutions, the coefficient of the rotational energy, $\hbar^2/2\Im_{\rm rigid}$, at each iteration is shown in Fig. \ref{Fig2_MOInodam}.
Here the moment of inertia is evaluated by the rigid body formula $\Im_{\rm rigid}=m_N\langle x^2+z^2\rangle$,
and the figure shows the value of $\hbar^2/2\Im_{\rm rigid}$ as a function of iteration numbers at rotational frequencies $\hbar\omega=0.5$, 1.0, ..., 4.0 MeV.
The iterations are terminated at the 10000-th iteration.
It is seen that the results of $\hbar^2/2\Im_{\rm rigid}$ in the 500-th iteration are very close to those in the 10000-th iteration.

According to the final values of $\hbar^2/2\Im_{\rm rigid}$, one can classify the final states into three groups:
(a) ground state for $\hbar\omega=0.5$, $1.0$ and $1.5$ MeV, (b) LCSs for $\hbar\omega=2.0$ and $2.5$ MeV,
and (c) fission for $\hbar\omega=3.0$, $3.5$ and $4.0$ MeV.
At rotational frequencies $\hbar\omega=0.5$, $1.0$ and $1.5$ MeV, LCS is not stable against the bending motion,
and finally it turns into the ground state.
With the increasing rotational frequency (2.0 and 2.5 MeV), the strong centrifugal force stabilizes LCSs against the bending motion.
The role of rotation for the stabilization of LCSs has been investigated in $^{16}$O and $^{24}$Mg by nonrelativistic DFT calculations \cite{Ichikawa2011O16LinearChain, Iwata2015PRC_LCS_Mg},
and similar conclusions have been found there.

\begin{figure}[h!]
  \centering
  \includegraphics[width=0.5\textwidth]{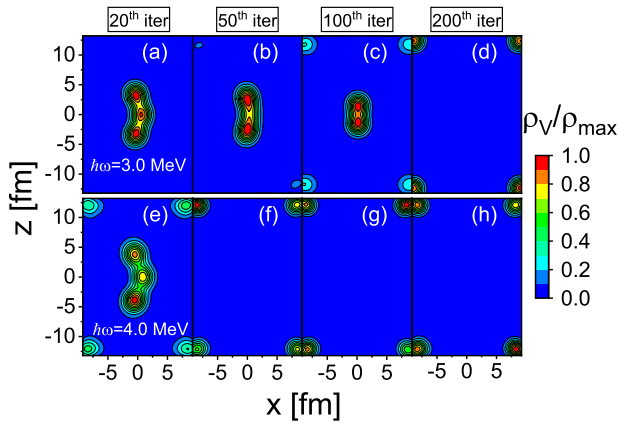}\\
  \caption{Total nucleon density distributions in the $x$-$z$ plane ($y$ direction is integrated) for the 20-th, 50-th, 100-th,
  and 200-th iterations at rotational frequency $\hbar\omega$ = 3.0 (upper panels) and 4.0 MeV (lower panels).
  In each figure, the density is normalized with respect to its maximum density $\rho_{\rm max}$. }\label{Fig3_Densnodam}
\end{figure}
However, in previous nonrelativistic DFT calculations, the fissions at high rotational frequencies have not been discussed.
In the following, a detailed analysis for these fissions will be performed by examining the density distributions and single-particle levels.

Taking $\hbar\omega=3.0$ and 4.0 MeV as examples,
the density distributions for the 20-th, 50-th, 100-th, and 200-th iterations are shown in Fig. \ref{Fig3_Densnodam}.
The fission processes show anomalies in density distributions.
In Figs. \ref{Fig3_Densnodam} (a)-(d) ($\hbar\omega=3.0$ MeV) and Figs. \ref{Fig3_Densnodam} (e)-(h) ($\hbar\omega=4.0$ MeV),
the densities distribute mainly in the central part of and/or the edge of the box, and there is no visible density distribution in between.
After checking the density distributions for every iteration, one can find the same anomalies in the density distributions.

\begin{figure}[h!]
  \centering
  \includegraphics[width=0.5\textwidth]{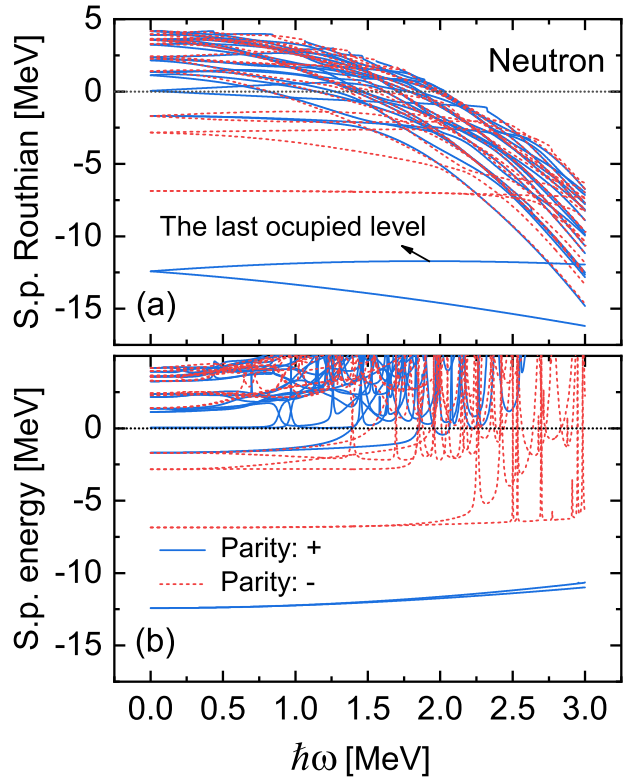}\\
  \caption{Neutron single-particle (s.p.) Routhians (a) and energies (b) as functions of $\hbar\omega$ with the potential frozen as the one for the LCS of $^{12}$C at $\hbar\omega=0.0$ MeV.
  The solid and dashed lines denote the positive and negative parity levels, respectively.
  }\label{Fig4_SpR_E}
\end{figure}
To understand these anomalies in the density distributions shown in Fig. \ref{Fig3_Densnodam}, we firstly calculate the LCS at $\hbar\omega=0.0$ MeV.
Although it is not a local minimum but rather a saddle point with respect to bending motion, we can still get a converged LCS self-consistently with three initial $\alpha$s on a straight line \cite{Maruhn2010NPA_LCS_Carbon}.
We freeze the potential at $\hbar\omega=0.0$ MeV and change the $\hbar\omega$ from 0.0 to 3.0 MeV.
In Fig. \ref{Fig4_SpR_E}, the obtained single-particle Routhians and energies for the neutrons are shown as functions of $\hbar\omega$.
As seen in Fig. \ref{Fig4_SpR_E} (a), some positive single-particle Routhians at $\hbar\omega=0.0$ MeV go down drastically and even cross with the occupied ones with increasing rotational frequency.
However, their single-particle energies are positive as shown in fig. \ref{Fig4_SpR_E} (b).
The steep slopes of these levels as displayed in Fig. \ref{Fig4_SpR_E} (a) mean that they have extreme large angular momenta.
The density distributions for these levels are mainly near the edge of the box instead of the central part.
Obviously, they are unphysical continuum.
Therefore, we can conclude that the fissions shown in Fig. \ref{Fig3_Densnodam} arise from the occupation of the unphysical continuum.

\begin{figure}[h!]
  \centering
  \includegraphics[width=0.5\textwidth]{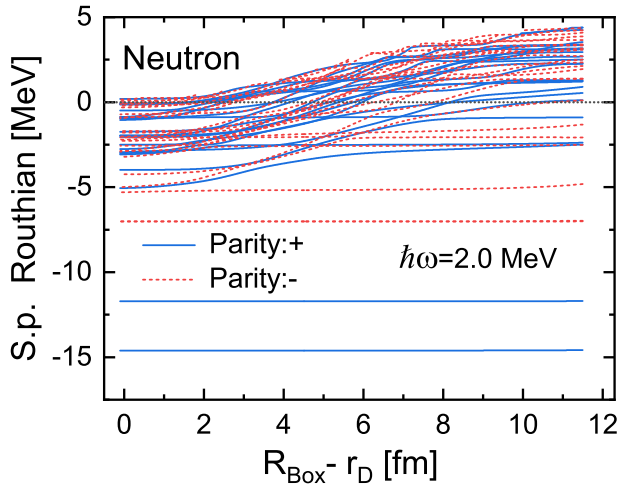}\\
  \caption{With the potential frozen as the one for the LCS of $^{12}$C at $\hbar\omega=0.0$ MeV, neutron single-particle Routhians with the damped cranking term as a function of $R_{\textrm{Box}}-r_D$.
  The $R_{\textrm{Box}}$ is defined as, $R_{\textrm{Box}}=[(L_x/2)^2+(L_y/2)^2+(L_z/2)^2]^{1/2}\approx18.5$ fm.
  The damping parameter $a_D=0.2$ fm.
  The conventions of the lines are the same as those in Fig. \ref{Fig4_SpR_E}.}\label{Fig5_SpR_Rd}
\end{figure}
We note, in the H.O. basis calculations, the unphysical continuum is excluded due to the artificial barrier of H.O. potential.
Here in the 3D lattice space calculations, however, the unphysical continuum appears and should be removed.
For this purpose, a Fermi-type damping function,
\begin{equation}
  f_D(r)=\frac{1}{1+e^{(r-r_D)/a_D}},
\end{equation}
is introduced for the cranking term $-\omega \hat{j}_y$ to exclude the unphysical continuum,
where $r_D$ is an effective cut-off parameter and $a_D$ is a smoothing parameter.
It means that the cranking term $-\omega \hat{j}_y$ in Eq. \eqref{Dirac_equation} is replaced by a damped one $-\omega[f_D(r)\hat{j}_yf_D(r)]$.
With the potential frozen as the one for the LCS in $^{12}$C at $\hbar\omega=0.0$ MeV, the neutron single-particle Routhians as functions of $R_{\textrm{Box}}-r_D$ at $\hbar\omega=2.0$ MeV are shown in Fig. \ref{Fig5_SpR_Rd}.
Here $R_{\textrm{Box}}$ is defined as, $R_{\textrm{Box}}=[(L_x/2)^2+(L_y/2)^2+(L_z/2)^2]^{1/2}\approx18.5$ fm.
Decreasing $r_D$ or increasing $R_{\textrm{Box}}-r_D$, the negative single-particle Routhians at $\hbar\omega=0.0$ MeV stay almost constant,
whereas the single-particle Routhians in the unphysical continuum significantly increase.
Therefore, by choosing a suitable $r_D$, the effects of unphysical continuum can be removed quite nicely,
and the influence on bound levels is negligible.
The values of $r_D=9$ fm and $a_D=0.2$ fm are adopted in the following calculations if not specified.

\begin{figure}[h!]
  \centering
  \includegraphics[width=0.5\textwidth]{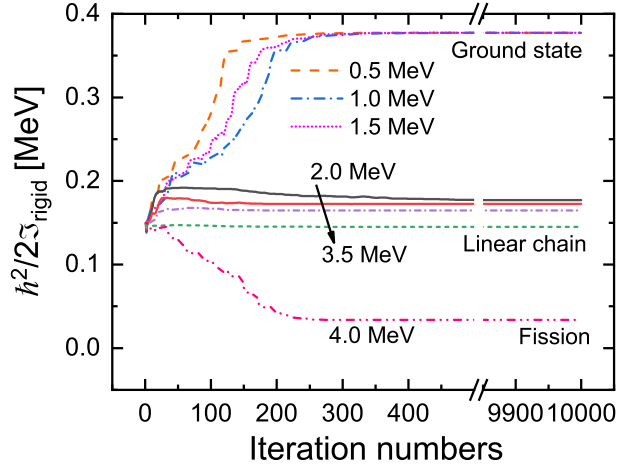}\\
  \caption{Same as in Fig. \ref{Fig2_MOInodam} but with the cranking term replaced by the damped one $-\omega f_D(r)\hat{j}_yf_D(r)$.}\label{Fig6_MOIdam}
\end{figure}
Then we perform the same calculations as in Fig. \ref{Fig2_MOInodam} but with the cranking term replaced by the damped one $-\omega f_D(r)\hat{j}_yf_D(r)$.
The coefficient of the rotational energy, $\hbar^2/2\Im_{\rm rigid}$, is shown in Fig. \ref{Fig6_MOIdam}
as a function of the iteration number.
For $\hbar\omega\leq2.5$ MeV, it is seen that the final results in Figs. \ref{Fig2_MOInodam} and \ref{Fig6_MOIdam} are identical.
In Fig. \ref{Fig6_MOIdam}, one can also find that the LCSs are stable against fission even
at $\hbar\omega=3.0$ and $3.5$ MeV, and fission finally occurs at $\hbar\omega=4.0$ MeV.

\begin{figure}[h!]
  \centering
  \includegraphics[width=0.5\textwidth]{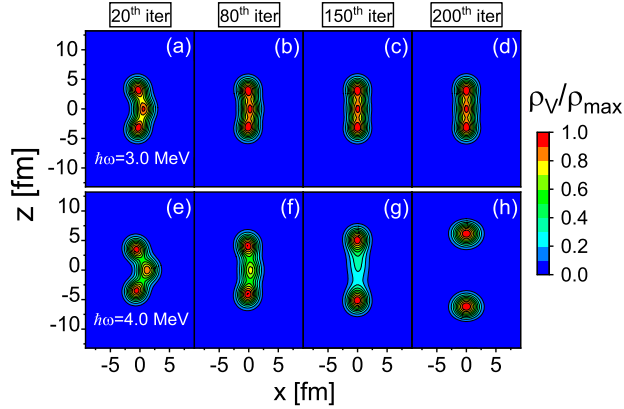}\\
  \caption{Same as in Fig. \ref{Fig3_Densnodam} but with the cranking term replaced by the damped one $-\omega f_D(r)\hat{j}_yf_D(r)$.}\label{Fig7_Densdam}
\end{figure}
After adopting the damped cranking term, the density evolutions at $\hbar\omega=3.0$ MeV and 4.0 MeV are shown in Fig. \ref{Fig7_Densdam}.
In the upper panels, Figs. \ref{Fig7_Densdam} (a)-(d) ($\hbar\omega=3.0$ MeV) show the transition of the LCSs from twist to straight with iteration.
As seen in Figs. \ref{Fig7_Densdam} (c) and (d),
since the LCSs are well confined within 9 fm,
the choice of the damping parameter $r_D=9$ fm is suitable.
In Figs. \ref{Fig7_Densdam} (e)-(h) ($\hbar\omega=4.0$ MeV), the fission process behaves like a liquid drop,
and this phenomenon coincides with the general understanding on fission in nuclear physics.
Therefore, we can conclude that, although the ``fissions'' in Figs. \ref{Fig2_MOInodam} and \ref{Fig3_Densnodam} arise from the occupation of unphysical continuum, the fission does occur at high frequencies as shown in Figs. \ref{Fig7_Densdam} (e)-(h).

\begin{figure}[h!]
  \centering
  \includegraphics[width=0.5\textwidth]{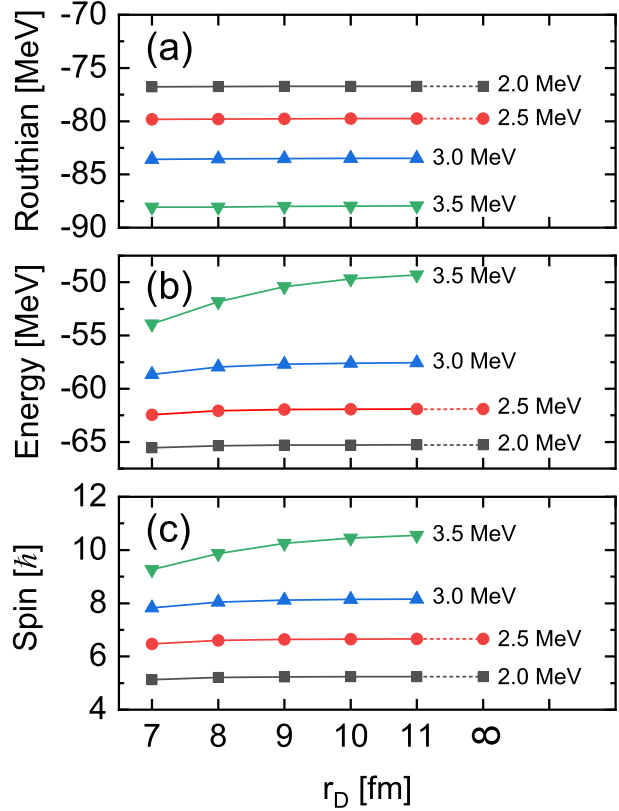}\\
  \caption{Total Routhian, energy, and spin versus the damping parameter $r_D$ at $\hbar\omega=$2.0, 2.5, 3.0, 3.5 MeV.
  The results of $r_D=\infty$ correspond to the calculations without the damping function.}\label{Fig8_REI_Rdam}
\end{figure}

In Fig. \ref{Fig8_REI_Rdam}, we show the total Routhians, energies, and spins versus the damping parameter $r_D$ at $\hbar\omega=$2.0, 2.5, 3.0 and 3.5 MeV.
The results of LCS at $\hbar\omega=2.0$ and 2.5 MeV without the damping function are also shown,
denoted as $r_D=\infty$.
For comparisons, the energies of the ground state and the LCS at $\hbar\omega=0.0$ MeV are $-87.8$ MeV and $-71.3$ MeV, respectively.
The obtained ground-state energy is in a good agreement with the datum $-92.2$ MeV \cite{AME2016}.
In Fig. \ref{Fig8_REI_Rdam} (a),
it can be found that the total Routhian remains nearly constant with $r_D$,
whereas the total energy and spin depend on $r_D$ as shown in Figs. \ref{Fig8_REI_Rdam} (b) and (c).
To achieve the convergence for total energy and spin,
the higher rotational frequency the larger $r_D$ is required.
The change of the total energies with $r_D=9$ fm and 11 fm are 0.2\% at  $\hbar\omega=3.0$ MeV and 2.2\% at $\hbar\omega=3.5$ MeV, respectively.
Therefore, for the present calculations, $r_D=9$ fm is a reasonable choice.

\begin{figure}[h!]
  \centering
  \includegraphics[width=0.5\textwidth]{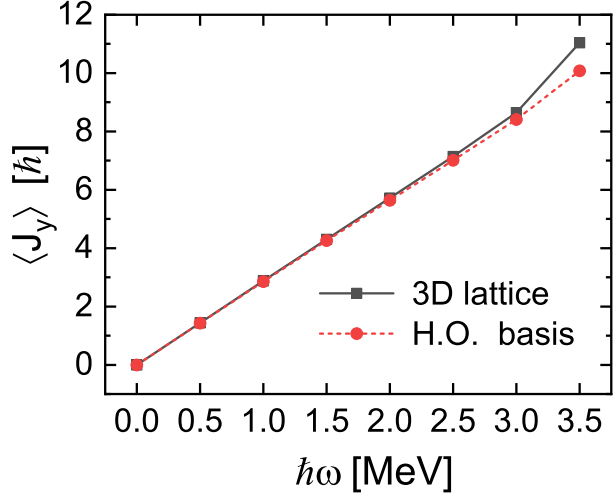}\\
  \caption{Angular momentum for the LCSs of $^{12}$C as a function of rotational frequency $\hbar\omega$.
  The solid line denotes the results of 3D lattice calculations with damping parameters $r_D=11.0$ fm and $a_D=0.2$ fm.
  The dashed line shows the results of harmonic oscillator (H.O.) basis expansion method with 12 major shells \cite{Zhao2015Rod-shaped}.}\label{Fig9_MOI}
\end{figure}
Finally, we show the angular momenta for the LCSs of $^{12}$C as a function of rotational frequency $\hbar\omega$ in Fig. \ref{Fig9_MOI}.
To get the converged LCSs at the lower rotational frequencies,
all 3D lattice calculations start with three $\alpha$s on a straight line.
The damping parameters $r_D=11.0$ fm and $a_D=0.2$ fm are adopted in the 3D lattice calculations.
For comparison, the results given by the H.O. basis expansion method with 12 major shells are also presented \cite{Zhao2015Rod-shaped}, where the reflection symmetry is imposed.
In Fig. \ref{Fig9_MOI}, except at $\hbar\omega=3.5$ MeV,
the angular momentum $\langle J_y\rangle$ almost increases with $\hbar\omega$ linearly.
It reveals that the moments of inertia (MOIs) $\Im$ are nearly constant below $\hbar\omega=3.0$ MeV.
The results of 3D lattice and H.O. basis calculations are almost identical up to $\hbar\omega$=3.0 MeV.
In comparison with Fig. \ref{Fig8_REI_Rdam} (c), the deviation at $\hbar\omega$=3.5 MeV may be attributed to the smaller model space adopted in the H.O. basis calculations compared to the 3D lattice calculations.
Ignoring results at $\hbar\omega=3.5$ MeV, the MOIs $\Im$ can be obtained by fitting $\langle J_y\rangle$ with $\hbar\omega$ linearly,
and they are 2.87 $(\textrm{MeV})^{-1}\hbar^2$ and 2.81 $(\textrm{MeV})^{-1}\hbar^2$ for 3D lattice and H.O. basis calculations, respectively.
The corresponding coefficients of the rotational energy, $\hbar^2/2\Im$, are evaluated as 0.174 MeV and 0.178 MeV respectively, and they are very close to each other.
Although there are many efforts to search the LCSs in $^{12}$C, no firm evidence has been found in experiments.
To stabilize three-$\alpha$ LCSs, adding valence neutrons has been suggested \cite{Itagaki2001MolecularOrbit, Maruhn2010NPA_LCS_Carbon, Zhao2015Rod-shaped}.
By adding two valence neutrons, the evidence for the existence of LCSs in $^{14}$C has been reported by several experimental groups.
The rotational bands with large MOI $\hbar^2/2\Im=0.12$ MeV \cite{Oertzenn2004C14_LCS, Freer2014C14_LCS_C14} and $\hbar^2/2\Im=0.19$ MeV \cite{Yamaguchi2017C14LCS} have been found in $^{14}$C,
which is a signal for the existence of LCSs.
In Ref. \cite{Zhao2015Rod-shaped}, it is found that MOIs of the LCSs in $^{12,14}$C are very close, so the MOIs of LCSs in $^{14}$C can be estimated by the results of $^{12}$C.
As we can see, $\hbar^2/2\Im=0.174$ MeV in this work is very close to the $\hbar^2/2\Im=0.19$ MeV in Ref. \cite{Yamaguchi2017C14LCS},
whereas it is larger than $\hbar^2/2\Im=0.12$ MeV in Refs. \cite{Oertzenn2004C14_LCS, Freer2014C14_LCS_C14}.
One might consider the experimental values $\hbar^2/2\Im=0.12$ and $0.19$ MeV correspond to different configurations,
and further investigations are required.

\section{SUMMARY}\label{summary}
The equation of motion in cranking covariant density functional theory is solved on a 3D lattice with the inverse Hamiltonian and the Fourier spectral methods.
The cranking CDFT in 3D lattice space is then applied to study the stability of the LCSs in $^{12}$C against the bending and fission at various rotational frequencies.

For cranking CDFT calculations in 3D lattice space, the single-particle Routhians of the unphysical continuum with large angular momenta go down drastically with rotational frequency and even cross with the occupied one.
This leads to fissions and the anomalies in the density distributions, namely, the densities distribute mainly in the central part of and/or the edge of the box, and there is no visible density distribution in between.

To exclude the unphysical continuum and avoid the anomalies in the density distributions,
a Fermi-type damping function is introduced for the cranking term.
After adopting the damped cranking term with reasonable damping parameters,
it is found that the linear chain structures are stable with the rotational frequency $\hbar\omega$ in the range of $\sim$2.0 MeV to $\sim$3.5 MeV.
The lower rotational frequency gives the ground state, while the higher one leads to the fission.
The moments of inertia for the rotational band of the linear chain states in $^{12}$C obtained are compared with the experimental ones in $^{14}$C,
and a good agreement is found.

\begin{acknowledgments}
Z.X.R thanks F.Q. Chen for helpful discussions.
This work was supported in part by the National Key R\&D Program of China (Contract No. 2018YFA0404400),
the National Natural Science Foundation of China (Grants No. 11335002 and No. 11621131001),
and the Laboratory Computing Resource Center at Argonne National Laboratory.
\end{acknowledgments}

\end{document}